\documentstyle[12pt,epsfig]{article}                                                                                                                        \parskip 2mm plus 2mm minus 2mm                                                            \newlength{\dinwidth}                                                            \newlength{\dinmargin}                                                                                                                        \textheight23.0cm \textwidth17.0cm                                                                                                                                                                                                                                                \oddsidemargin -1.0in                                                                                                                                                                                                                                                \marginparsep 8pt \marginparpush 5pt                                                            \topmargin -42pt                                                            \headheight 12pt                                                            \headsep 30pt \footheight 12pt \footskip 24pt                                                                                                             
\def\lapproxeq{\lower .7ex\hbox{$\;\stackrel{\textstyle                                                            <}{\sim}\;$}}                                                            \def\gapproxeq{\lower .7ex\hbox{$\;\stackrel{\textstyle                                                            >}{\sim}\;$}}
                                                                                          \def\be{\begin{equation}}                                                            \def\ee{\end{equation}}                                                            \def\bea{\begin{eqnarray}}                                                            \def\eea{\end{eqnarray}}                                                                                                                        \def\funp{{I\!\!P}}

\begin{document}                                                            \titlepage
\begin{flushright}                                                            IPPP/01/13 \\
DCPT/01/26 \\                                                            27 March 2001 \\                                                            \end{flushright}                                                                                     
\vspace*{2cm}                                                                       
\begin{center}{\Large \bf Summary Talk:}\\   

\vspace*{1cm}
{\Large \bf First Workshop on Forward Physics and} \\

\vspace*{0.4cm}
\renewcommand{\thefootnote}{\fnsymbol{footnote}}                                                            {\Large \bf Luminosity Determination at the LHC\footnote[2]{Helsinki, 31 October--3
November, 2000}} \\

\vspace*{1cm}                                                            A.D. Martin$^a$ \\                                                                       
\vspace*{0.5cm}                                                            $^a$ Department of Physics and Institute for Particle Physics Phenomenology, University of 
Durham, Durham, DH1 3LE 
\end{center}                                                                        
\vspace*{2cm}                                                                       
\begin{abstract}                                                            An attempt is made to summarize the discussion at the Workshop, except for the panel discussion on
the ability of the LHC detectors to accommodate forward reactions.  The Workshop focused on two main
topics.  The first topic was forward physics at the LHC.  Predictions were made for forward reactions,
including elastic scattering and \lq soft\rq\ diffractive processes, in terms of (multi)
Pomeron exchange, using knowledge gained at lower energies.  The survival probability of rapidity gaps
accompanying hard subprocesses was studied.  The nature of the Pomeron, before and after QCD, was
exposed, and some aspects of small $x$ physics at the LHC were considered.  The second topic of the
Workshop concerned the accuracy of the luminosity measuring processes at the LHC.  Attention concentrated on three methods.  The classic approach based on the optical theorem, secondly, the observation of the pure QED process of lepton-pair ($\ell^+ \ell^-$) production by photon-photon fusion and, finally, the measurement of inclusive $W$ and $Z$ production.
\end{abstract}

\newpage
\section{Introduction}

A lot of effort has justifiably been spent on the central detectors of the LHC experiments so
that \lq hard\rq\ interactions can be triggered on, and observed, in order to expose New
Physics.  However the vast majority of interactions are \lq soft\rq\ with particles
predominantly going forward, and the physics of this domain is one of the subjects of the
Workshop.  The relevant processes, forward elastic scattering and \lq soft\rq\ diffraction, are
driven by Pomeron (or rather by multi-Pomeron) exchange.  They are interesting in their own
right, although in the period after the advent of QCD they did not attract so much attention\footnote{An exception is Bjorken \cite{BJ1} who continued to emphasize the importance of experiments to detect complete events, including those in the forward region.}.
However in the last few years the situation has changed.  A stimulus came from the
observation of diffractive processes at HERA and the Tevatron, characterised by the presence
of rapidity gaps.  Moreover, diffractive processes have been proposed as possible ways to
identify New Physics.  For example, the central production of a Higgs boson with a rapidity
gap on either side is advocated as a possible discovery channel at the LHC.  The chance that
these gaps survive the soft rescattering of the colliding hadrons was one of the topics of
discussion.

The second subject of the Workshop was the way to measure the luminosity of the LHC.  
Three methods were discussed: 
\begin{itemize}
\item[(i)] the classic method based on the optical theorem,
\be
\label{eq:a1}
\left . \frac{d \sigma_{\rm el}}{dt} \right |_{t = 0} \; = \; \frac{\sigma_{\rm tot}^2}{16\pi} \:
(1 + \rho^2),
\ee
where $\rho$ is the ratio of the real to the imaginary part of the forward elastic amplitude
(Coulomb effects have been neglected in (\ref{eq:a1})),

\item[(ii)] to measure pure QED $e^+ e^-$ or $\mu^+ \mu^-$ production via photon-photon
fusion
\be
\label{eq:a2}
pp \; \rightarrow \; p \: + \: \ell^+ \ell^- \: + \: p,
\ee

\item[(iii)] to measure inclusive $W$ or $Z$ production.
\end{itemize}
At first sight it might appear that an accurate measurement of the luminosity is not essential.
However, for example, precision measurements in the Higgs sector of accuracy of about 7\%
require the uncertainty in the measurement of the luminosity to be 5\% or less \cite{GA}.  At
this Workshop, Tapprogge \cite{TAPP} summarised the physics reasons why a precise
measurement of the luminosity of the LHC is important.

\section{Proposed forward $pp$ measurements}

Bozzo \cite{BOZ} presented the TOTEM experimental programme, which will take place in 
the very first runs at the LHC.  The plan is to measure $\sigma_{\rm tot}$ by a  
luminosity-independent method
\be
\label{eq:a3}
\sigma_{\rm tot} \; = \; \frac{16 \pi}{(1 + \rho^2)} \; \frac{(dN_{\rm el}/dt)_{t = 0}}{N_{\rm
el} + N_{\rm inel}},
\ee
with an absolute error of about 1~mb.  The TOTEM inelastic detector will be installed inside
the CMS experiment, with the elastic scattering \lq\lq roman pot\rq\rq\ detectors located at
distances in the interval 100 to 200~m from the crossing point.  The TOTEM measurements
need special runs at the \lq low\rq\ initial luminosity, with high $\beta^*$ optics for an accurate
measurement of the small scattering angles.  The detector should be efficient to within 2~mm of the
beam and can reach down to about $-t = 0.01~{\rm GeV}^2$.  They will collect about 100 
events/sec for $-t < 1~{\rm GeV}^2$ at a luminosity ${\cal L} = 10^{28}~{\rm cm}^{-
2}~{\rm s}^{-1}$.  The differential cross section will also be measured in the interval $1 < |t|
< 10~{\rm GeV}^2$ in the high luminosity runs.  TOTEM also plans an inclusive trigger for
the measurement of single diffraction.

We also heard at the Workshop about the novel microstation concept for forward
measurements \cite{NOM}.  The microstation is a light, compact device which could be
integrated with the beam pipe.  It could be used in situations where there are severe space and
mass limitations for the inelastic detector, such as in the ATLAS experiment.  The elastic
proton measurement would be like that for TOTEM, so again it should be possible to reach
down to $-t \sim 0.01~{\rm GeV}^2$.

Piotrzkowski \cite{PI} considered the possibility of using the LHC as a $\gamma\gamma$ collider,
and noted how the present TOTEM layout may be modified to allow the tagging of protons with small
energy losses.  In principle, this would allow the study of $\gamma\gamma \rightarrow H$ etc.\ in
a broad region of $\gamma\gamma$ centre-of-mass energy about 200~GeV, with an effective $\gamma\gamma$ luminosity that is reduced by about $10^{-3}$--$10^{-2}$ of that of the LHC.

Guryn \cite{GUR} presented the PP2PP experimental physics programme at RHIC.  Their
main goal is to make a detailed study of the spin dependence of the proton-proton interaction
in the kinematic range
\be
\label{eq:a4}
50 \; < \; \sqrt{s} \; < \; 500~{\rm GeV} \quad\quad\quad {\rm and} \quad\quad\quad 4 \;
\times \; 10^{-4} \; < \; |t| \; < \; 1.5~{\rm GeV}^2,
\ee
in order to probe features of the Pomeron.  At the beginning they will measure $d
\sigma_{\rm el}/dt$ down to $-t = 0.006~{\rm GeV}^2$, and study the $t$ dependence of the
slope.  Later they will explore the Coulomb-nuclear interference region, and determine 
$\rho$.  By measuring the spin asymmetries $A_N, A_{NN}, A_{LL}$ they will obtain
information on the five independent $pp \rightarrow pp$ helicity amplitudes.

Block \cite{BLOCK} discussed the determination of $\sigma_{\rm tot} (pp)$ at $\sqrt{s} \simeq 30$~TeV
from cosmic ray data.  He emphasized the importance of the relation between the slope $B$ of the elastic scattering distribution and $\sigma_{\rm tot}$ for energies well above the accelerator regime, and the reliance on a model of proton-air interactions.

\section{Extrapolation of $d\sigma_{\rm el}/dt$ to $t = 0$}

Vorobyov \cite{VOR} reminded us how at ISR energies, and below, it was possible to make
precision measurements of the elastic differential cross section at small momentum transfer
down into the Coulomb-nuclear interference region.  From these experiments one could make
a determination of the local slope
\be
\label{eq:a5}
B (t) \; = \; \frac{d \ln (d \sigma_{\rm el}/dt)}{dt}
\ee
as a function of $t$, measure the ratio $\rho$ of the real to imaginary part of the forward
amplitude and reliably determine the total cross section from the optical point.  However the
application of this method becomes more and more difficult as we go up to Tevatron and then
LHC energies, due to the decrease of the typical scattering angle and the inability to 
experimentally reach the Coulomb-nuclear interference region.

\begin{figure}[p]
\begin{center}\vspace{-1cm}\epsfig{figure=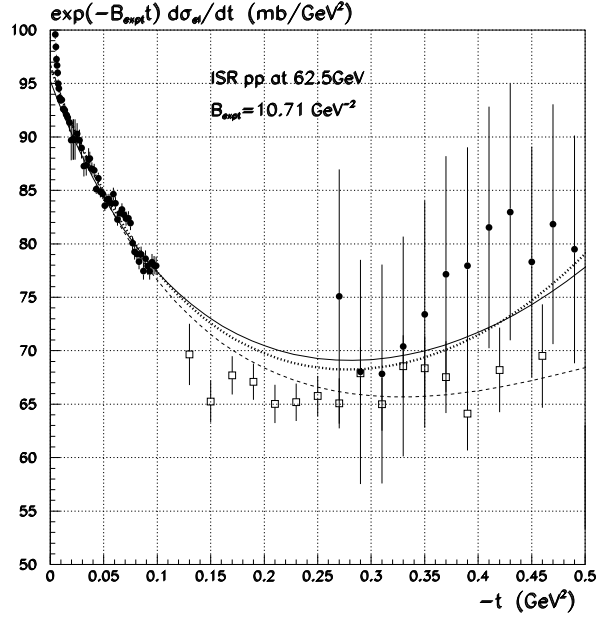,height=3.5in}
\caption{ISR data for $pp\;d \sigma_{\rm el}/dt$, with the experimental exponential
form divided out, compared with the description given by a multi-Pomeron analysis
\cite{KMR}.  The Coulomb-nuclear interference is evident in the data at very small $t$.  The
dashed curve should be disregarded, as it shows the prediction with high-mass diffraction
neglected.  The continuous and dotted curves are obtained using two extremum models for high-mass diffraction.\vspace{-1.5cm}\label{fig:fig1}}
\end{center}
\end{figure}

\begin{figure}[p]
\begin{center} \epsfig{figure=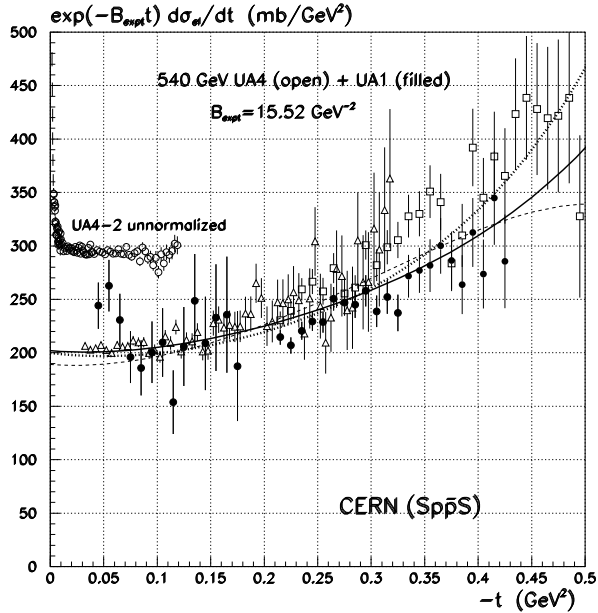,height=3.5in}
\caption{As for Fig.~1 but showing $Sp\bar{p}S$ elastic data.  The most recent UA4
data are unnormalised and are plotted higher for clarity.
\label{fig:fig2}}
\end{center}
\end{figure}

\begin{figure}[p]
\begin{center} \epsfig{figure=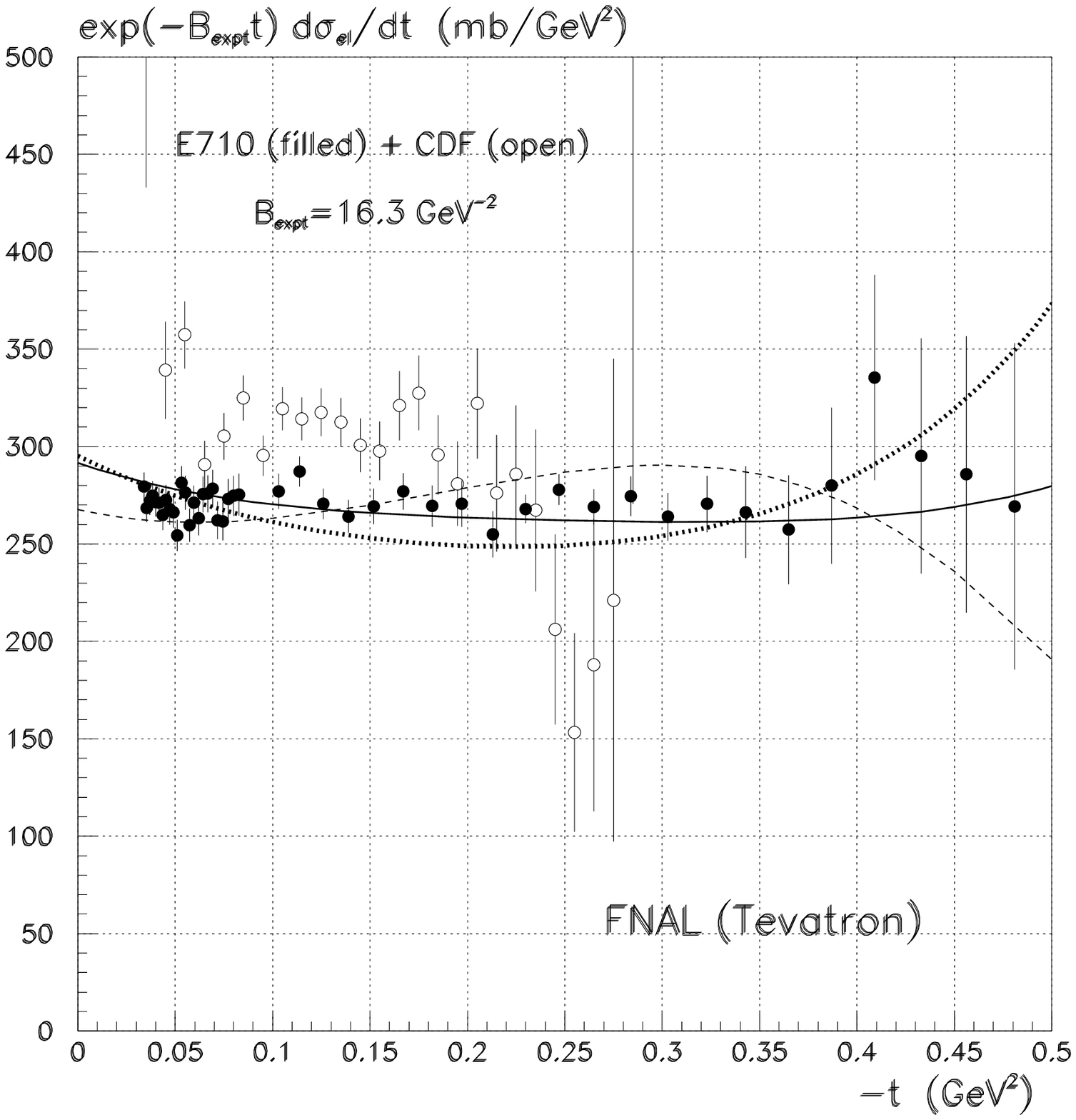,height=3.3in}
\caption{As for Fig.~1 but showing Tevatron elastic data. \label{fig:fig3}}
\end{center}
\end{figure}

Figs.~1, 2 and 3 show data at ISR, CERN $Sp\bar{p}S$ and Tevatron energies respectively.
In the first two plots we see the Coulomb interference spike at very small $t$, whereas at the
Tevatron the spike is no longer accessible and the data of the two experiments extrapolate to
different values at $t = 0$.  Moreover at the ISR we see, from Fig.~1, a change of the local slope with $t$,
\be
\label{eq:a6}
B (0) \: - \: B ( | t | = 0.2~{\rm GeV}^2) \; \simeq \; 2~{\rm GeV}^{-2},
\ee
whereas at the Tevatron the data suggest less variation for
$| t | \lapproxeq 0.4~{\rm GeV}^2$, see Fig.~3.  A global description of these and other forward data gives the $t$ dependence of
the local slopes shown in Fig.~4 \cite{KMR}.  The dashed curves should be ignored as they come from a description which does not include high-mass diffraction.  The remaining two curves (continuous and dotted) are obtained using two extremum models for high-mass diffraction, and give a measure of the level of the theoretical ambiguity in $B (t)$.  The predictions at the LHC energy mean that
we should be able to extrapolate the TOTEM measurements to $t = 0$ with an error of less
than 0.5\% coming from the variation of the local slope.  Even if the cross section were
measured only in the interval $0.05 < |t| < 0.15~{\rm GeV}^2$, then the uncertainty at $t = 0$
would be less than 3\% due to the variation of $B (t)$.

\begin{figure}[p]
\begin{center}
\epsfig{figure=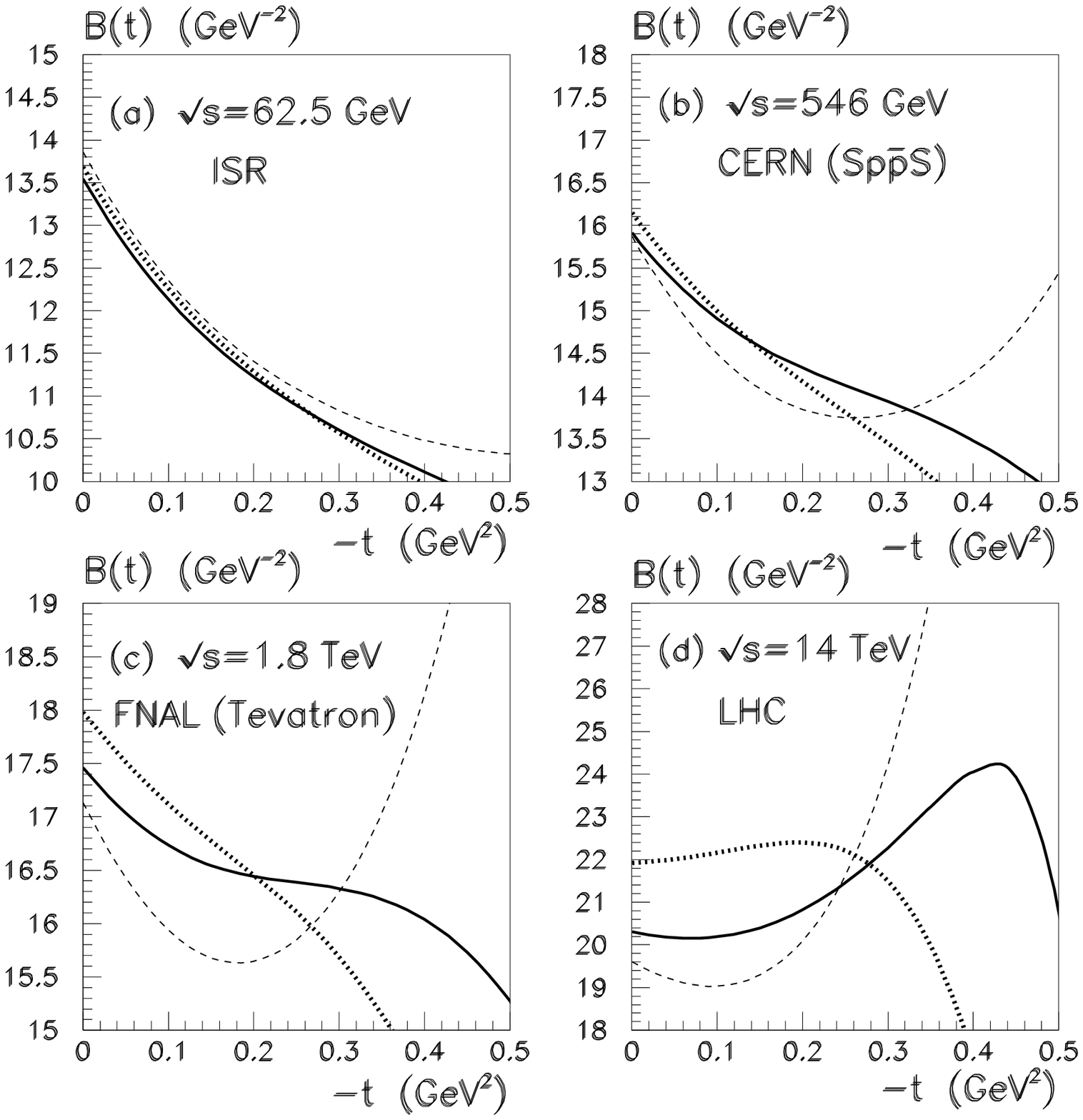,height=3.4in}
\caption{The multi-Pomeron model \cite{KMR} predictions for the $t$-dependence of
the local elastic slope, $B (t)$ of (\ref{eq:a5}), at ISR, $Sp\bar{p}S$, Tevatron and LHC
energies.  The rise of the dashed curve is due to the diffractive minimum, which is considerably
modified by the inclusion of high-mass diffraction, as shown by the continuous and dotted curves.
\label{fig:fig4}}
\end{center}
\end{figure}

\section{The Pomeron and the description of forward data}

High energy $pp$ scattering at small momentum transfer $t$ is driven by Pomeron
$t$-channel exchange.  In fact we were reminded at the Workshop \cite{LIP,MAR} how in 
1960, Gribov startled the community by showing that the behaviour of the $pp$ amplitude
\be
\label{eq:a7}
A (s, t) \; = \; i s f (t),
\ee
which gives the asymptotic behaviour\footnote{The $s-u$ crossing invariance is an important
ingredient in the second equality in (\ref{eq:a8}), the Pomeranchuk theorem.}
\be
\label{eq:a8}
\sigma_{\rm tot} \; = \; {\rm constant}, \quad\quad \sigma (pp) \; = \; \sigma (p\bar{p}),
\ee
contradicts $t$-channel unitarity.  Gribov noted that a possible solution was to introduce the
Regge behaviour
\be
\label{eq:a9}
A (s, t) \; = \; i s^{\alpha_\funp (t)} \: f (t),
\ee
with a Pomeron trajectory $\alpha_\funp (t) = \alpha_\funp (0) + \alpha^\prime t$, with
intercept $\alpha_\funp (0) = 1$ and slope $\alpha^\prime > 0$.  This is in agreement with
$t$-channel unitarity, while still giving relations (\ref{eq:a8}).  In addition to the Pomeron
pole, there are also Regge cuts coming from multiple Pomeron exchange, which led to the
development of Gribov's Reggeon calculus, involving renormalisation of the bare pole, 
Mandelstam crossed diagrams \cite{MAN}, AGK cutting rules \cite{AGK}, etc.  This \lq  
soft\rq\ Pomeron was the subject of the talks by Landshoff \cite{LAND} and Kaidalov
\cite{KAID}.

Landshoff \cite{LAND} showed that a good description of the available data for high energy
$\sigma_{\rm tot}$ and $d \sigma_{\rm el}/dt$ at small momentum transfer is given by a 
simple effective Pomeron pole trajectory
\be
\label{eq:a10}
\alpha_{\rm eff} (t) \; = \; 1.08 \: + \: \alpha^\prime t
\ee
with $\alpha^\prime = 0.25~{\rm GeV}^{-2}$.  This is a remarkable simplification, but of
course it is necessarily incomplete for the reasons that are mentioned below.
\begin{figure}[htb]
\begin{center}
\epsfig{figure=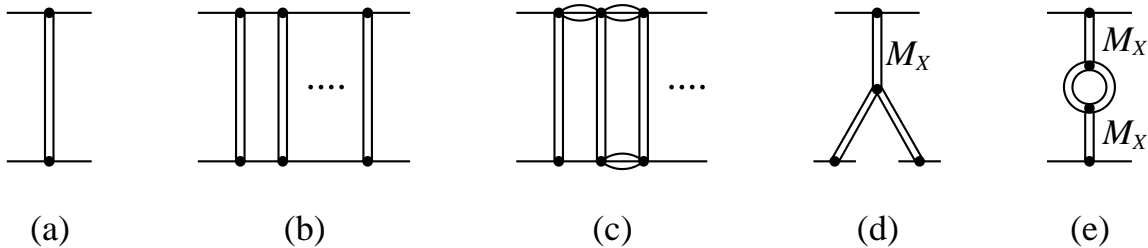,height=1.5in}
\caption{The Pomeron exchange contribution (a), together with unitarity corrections (b--e).  Note
that graphs (d, e) are the \lq square\rq\ of the single- and double-diffractive dissociation
amplitudes respectively --- and that their constituent Pomeron lines are subjected to 
rescattering corrections.
\label{fig:fig5}}
\end{center}
\end{figure}

Kaidalov \cite{KAID} explained that to describe phenomena in the forward region, it is 
important to extend single-Pomeron exchange so as to explicitly include the multi-Pomeron
effects.  At very high energy such effects are needed to restore $s$-channel unitarity, but, as
we will see in a moment, they are also necessary at current energies.  Fig.~5 shows examples
of typical corrections to the bare Pomeron pole of diagram (a).  First, iterations of the pole
amplitude via elastic unitarity gives contributions of the type shown in diagram (b).  If we
take into account the possibility of proton excitations $(p \rightarrow N^*)$ in intermediate
states, then we must include contributions such as that in diagram (c).  These iterations are
implemented in terms of a two-channel eikonal formalism \cite{TMR,ABK}.  Note that diagrams (b) and (c) are very
symbolic --- by implication they incorporate the Mandelstam crossed diagrams \cite{MAN}
and satisfy the AGK cutting rules \cite{AGK}.  The second $(N^*)$ channel effectively 
allows for {\it low mass} diffractive dissociation.  The excitation into {\it high mass}
$(M_X)$ states is described by the triple-Pomeron graph (d) for single-diffractive
dissociation (with cross section $\sigma_{\rm SD}$), and by (e) for double-diffractive 
dissociation $(\sigma_{\rm DD})$.  To be self-consistent, the Pomeron lines in (d) and (e)
actually represent the final Pomeron amplitude with all the screening effects included.  In
addition to (d), there is an equal contribution $\sigma_{\rm SD}$ from dissociation of the
lower proton only.  The contribution of graphs of the type (b)--(e) are not negligible.  Indeed,
using the AGK rules, it may be estimated that the correction to (a) is
\be
\label{eq:a11}
\sigma_D/\sigma_{\rm tot} \; \equiv \; (\sigma_{\rm el} \: + \: 2 \sigma_{\rm SD} \: + \: \sigma_{\rm DD})/\sigma_{\rm tot} \: \sim \: 0.4
\ee
at the LHC.
A most convincing way to see the necessity of multi-Pomeron rescattering effects at current
energies is to look at the energy behaviour of $\sigma_{\rm SD}$.  If only the bare Pomeron
is used, as in diagram (d), then the cross section grows as
\be
\label{eq:a12}
\sigma_{\rm SD} \; \sim \; s^{2 \alpha_\funp (t) - 2},
\ee
whereas when rescattering effects are included
\be
\label{eq:a13}
\sigma_{\rm SD} \; \sim \; \sigma_{\rm tot}/\ln s.
\ee
The difference is dramatic, as Fig.~6 shows.  The physical origin of this result is that the high
energy proton-proton interaction reaches the black-disc limit for central values of the impact
parameter, and forces inelastic diffraction to come from the peripheral region with $b \sim c
\ln s$.

\begin{figure}[htb]
\begin{center}
\epsfig{figure=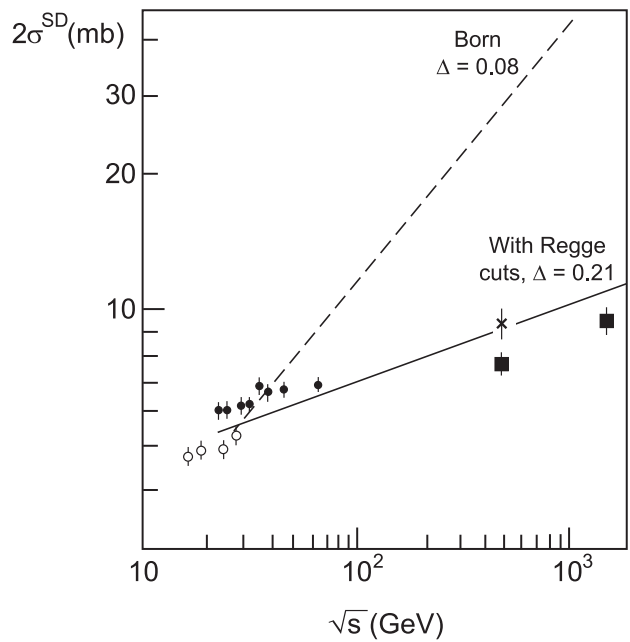,height=3.3in}
\caption{The energy behaviour of the cross section for single-diffractive dissociation 
compared to the predictions of the effective pole and multi-Pomeron approaches
\cite{KAID}.  $\Delta \equiv \alpha (0) - 1$.\label{fig:fig6}}
\end{center}
\end{figure}

A related observation was made by Schlein \cite{SCH}.  He showed that the single diffractive data in the ISR-Tevatron energy range can be described by the triple-Regge formula, provided that the effective intercept $\alpha_\funp (0)$ decreased with increasing $s$.  Such an $s$ dependent behaviour of the effective intercept may be interpreted as a manifestation of the multi-Pomeron exchange effects.The bare Pomeron, together with the multi-Pomeron corrections, should give a global
description of all forward phenomena, including $\sigma_{\rm tot}, d\sigma_{\rm el}/dt,
\rho, \sigma_{\rm SD}, \sigma_{\rm DD}$ and $\langle n_{\rm ch} \rangle$.  Several
analyses have been undertaken at various levels of sophistication to all or parts of the data.
Recent examples are given in \cite{KMR,BH,LEVIN}.  Perhaps the most complete study to 
date is that of Ref.~\cite{KMR}, which is in the spirit of the much earlier pioneering work of Kaidalov
et al.\ \cite{KPT}.  It describes the forward phenomena in high energy $pp$ (and $p\bar{p}$)
collisions using a multi-Pomeron approach which embodies:
\begin{itemize}
\item[(i)] {\it pion-loop} insertions in the bare Pomeron pole, which represent the nearest
singularity generated by $t$-channel unitarity,
\item[(ii)] a {\it two-channel eikonal} which incorporates the Pomeron cuts generated by
elastic and quasi-elastic (with $N^*$ intermediate states) $s$-channel unitarity.
\item[(iii)] high mass single and double {\it diffractive dissociation}. \end{itemize}
The resulting description of $d \sigma_{\rm el}/dt$ data is shown in Figs.~1--3 and the
corresponding $t$ dependence of the local slope $B (t)$ is given in Fig.~4.  Surprisingly, the
bare Pomeron pole parameters turn out to be
\be
\label{eq:14}
\Delta \; \equiv \; \alpha (0) \: - \: 1 \; \simeq \; 0.10, \quad\quad \alpha^\prime \; = \; 0,
\ee
which is to be contrasted to those of the effective pole (\ref{eq:a10}).  Thus the shrinkage of
the diffraction cone comes not from the bare pole, but rather has components from the three
ingredients, (i)--(iii), of the model.  That is, in the ISR-Tevatron energy range \cite{CET}\be
\label{eq:a15}
{\rm \lq\lq}\alpha_{\rm eff}^\prime" \; = \; (0.034 \: + \: 0.15 \: + \: 0.066)~{\rm GeV}^{-2}
\ee
from the $\pi$-loop, $s$-channel eikonalization and diffractive dissociation respectively.  We
saw \cite{VOR} that at lower energies the fixed target data require $\alpha_{\rm eff}^\prime
= 0.14~{\rm GeV}^{-2}$, which is consistent with (\ref{eq:a15}) since as the energy
decreases the effect of the eikonal and higher mass diffractive dissociation reduces. 
Moreover eikonal rescattering suppresses the growth of the cross section with $\sqrt{s}$, so
to describe the same data we require $\Delta > \Delta_{\rm eff} \simeq 0.08$.

\begin{figure}[htb]
\begin{center}
\epsfig{figure=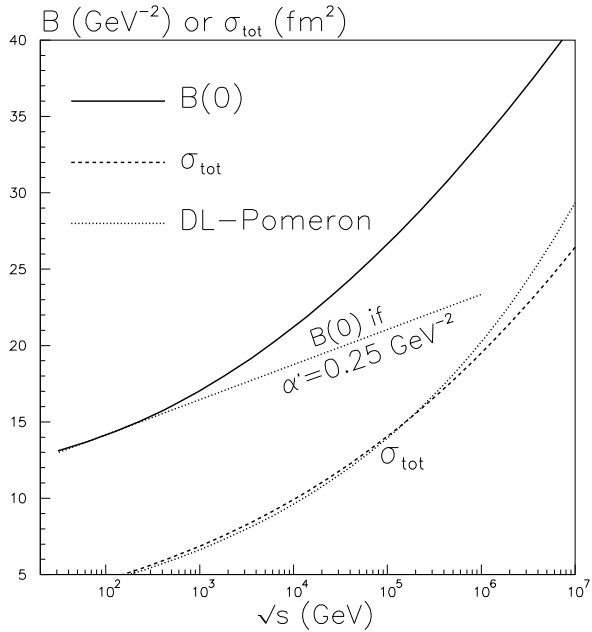,height=3.5in}
\caption{The energy behaviour of the $pp$ forward elastic slope and the total cross
section of the multi-Pomeron approach \cite{KMR}, compared with that obtained from the 
simple effective pole form of (\ref{eq:a10}).  $\sigma_{\rm tot}$ is plotted in fm$^2$ and $B
(0)$ in GeV$^{-2}$, since in these units the asymptotic black-disc limit takes, to a good
approximation, the simple form $B (0)/\sigma_{\rm tot} \rightarrow 1$.  The figure is taken from Ref.~\cite{CET}.\label{fig:fig7}}
\end{center}
\end{figure}

Another observable difference between the naive effective pole and the multi-Pomeron
approach is in the energy behaviour of the slope $B (0)$.  This is illustrated in Fig.~7,
together with the behaviour of $\sigma_{\rm tot}$.  It is seen that, while the two approaches
give similar values of $\sigma_{\rm tot}$ up to LHC energies, the predictions for the slope
$B (0)$ at the LHC already differ significantly.

In Table~1 we show some predictions for the forward observables at the LHC.  Note that the
BH predictions \cite{BH} are based on a single-channel eikonal, and do not therefore involve
inelastic diffraction.  The KMR prediction is an update of \cite{KMR}.

\begin{table}[htb]
\begin{center}
\begin{tabular}{|c|c|c|c|} \hline
& KPT \cite{KPT} & KMR \cite{KMR} & BH \cite{BH} \\
& 1986 & 2000 & 2000 \\ \hline
$\sigma_{\rm tot}$ (mb) & 103 & 108 & 107 \\
$2 \sigma_{\rm SD}$ (mb) & 12 & 15 & - \\
$\sigma_{\rm DD}$ (mb) & 13 & 8 & - \\
$B (0)$ (GeV$^{-2}$) & 21.5 & 20.7 & 19.4 \\
$\rho =$ Re/Im & 0.11 & 0.11 & 0.11 \\
$\sigma_{\rm el}$ (mb) & 26 & & 31 \\ \hline
\end{tabular}
\caption{Predictions for forward observables at the LHC.}
\end{center}
\end{table}

\section{Hard diffraction and gap survival probabilities}

There is much interest in the survival probability of rapidity gaps which feature in the various
hard diffractive and other high energy processes.  The rapidity gaps, which naturally occur
whenever we have colour-singlet $t$-channel exchange \cite{DKT,BJ1,BJ2}, tend to get populated by secondary particles from the soft rescattering processes.  A multi-Pomeron analysis incorporates rescattering in some detail and so allows the calculation of the survival probabilities $S^2$ of the gaps.  Recently attention has focused on the size of $S^2$, see for example \cite{GLM,KMR,BH,VAK,AR},
because of the possibility of extracting New Physics from hard processes, accompanied by
gaps, in an almost background-free environment and, from a theoretical viewpoint, because of
its reliance on subtle QCD techniques.  The theoretical calculations of $S^2$ do not allow for
the practical difficulties of isolating rapidity gap processes in the \lq pile-up\rq\ events which
may occur at high LHC luminosity.

To calculate $S^2$, it is convenient to work in impact parameter space.  Let ${\cal M} (s,b)$
be the amplitude of the particular hard process of interest at centre-of-mass energy
$\sqrt{s}$.  Then the probability that there is no extra inelastic interaction is
\be
\label{eq:a16}
S^2 \; = \; \frac{\int \: | {\cal M}|^2 \: e^{- \Omega} \: d^2 b}{\int \: | {\cal M} |^2 \: d^2 b},
\ee
where $\Omega (s,b)$ is the opacity (or optical density) of the interaction of the incoming
protons.  For simplicity we show the formula with the simple one-channel eikonal.  In practice the
numerator and denominator in (\ref{eq:a16}) are sums over the diffractive
eigenstates\footnote{States which diagonalize the diffractive part of the $T$ matrix and so
undergo only elastic scattering.}, each with their own characteristic opacity $\Omega_i$.  The
opacities $\Omega_i (s,b)$ reach a maximum at the centre of the proton and become small in
the periphery.  Clearly the survival probability $S^2$ depends strongly on the spatial
distribution of the constituents of the relevant subprocess, and on the dynamics of the whole
diffractive part of the scattering matrix.  Contrary to frequent claims in the literature, it is
important to note that $S^2$ is not universal, but depends on the particular hard subprocess,
as well as the kinematical configurations.  In particular, $S^2$ depends on the nature of the
colour-singlet exchange (Pomeron or, possibly, $W/Z$ or photon exchange) which generates
the gap \cite{KMR}, as well as on the characteristic momentum fractions carried by the active partons in
the colliding hadrons \cite{KKMR}.  This leads to a rich structure of the probability of rapidity gaps in
processes mediated by colour-singlet $t$-channel exchange.  The framework was introduced
long ago\footnote{Reviews can be found, for example, in Refs.~\cite{ABK,PUMP}.}
\cite{GW,GRIB}, but only with the advent of rapidity gap events being observed in hard 
processes at the Tevatron and at HERA, is this rich physics now revealing itself.  Clearly it is
important to have reliable estimates of the survival probabilities so as to be able to predict the
rate for rapidity gap processes at the LHC.  Indeed, the possibility of using central Higgs
production with a rapidity gap either side as a discovery channel at the LHC was discussed at
the Workshop by Khoze \cite{VAK}.

Examples of estimates of the survival probability for single- and double-diffractive
dissociation (SD, DD), and central diffraction (CD) are given in Table~2.  By central 
diffraction we mean a centrally produced state $X$ with rapidity gaps on either side.  The $b$
dependence of the various diffractive processes is
\be
\label{eq:a17}
| {\cal M} |^2 \; \propto \; \exp (-b^2/nB^\prime),
\ee
where $n = 3,4$ and 2 for SD, DD and CD respectively and $B^\prime$ is the slope of the
diffractive inclusive cross section \cite{KMR}.  The single channel eikonal model of
\cite{BH} gives values of $S^2$, very similar to those of the DD column.
\begin{table}[htb]
\begin{center}
\begin{tabular}{|c|c|ccc|} \hline
$\sqrt{s}$ & $B^\prime$ & \multicolumn{3}{|c|}{survival probability $S^2$ for:} \\
(TeV) & (GeV$^{-2}$) & SD & \quad\quad DD & \quad~ CD \\ \hline
\raisebox{-1.7ex}[0pt]{1.8} & 4.0 & 0.10 & \quad\quad 0.15 & \quad~ 0.05 \\
& 5.5 & 0.15 & \quad\quad 0.21 & \quad~ 0.08 \\ \hline
\raisebox{-1.7ex}[0pt]{14} & 4.0 & 0.06 & \quad\quad 0.10 & \quad~ 0.02 \\
& 5.5 & 0.09 & \quad\quad 0.15 & \quad~ 0.04 \\ \hline
\end{tabular}
\caption{The survival probability $S^2$ of rapidity gaps in single, double and central
diffractive processes at Tevatron and LHC energies \cite{KMR}.}
\end{center}
\end{table}

\begin{figure}[htb]
\begin{center}
\epsfig{figure=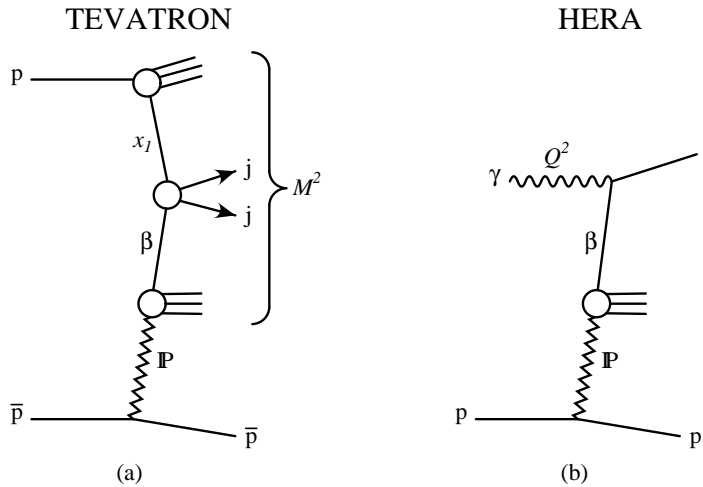,height=2.6in}
\caption{The partonic structure of (a) diffractive dijet production at the Tevatron and (b)
diffractive deep inelastic scattering at HERA.  Process (a) has soft rescattering corrections
(not shown) which give secondaries which populate the rapidity gap associated with Pomeron
exchange, and hence suppress the cross section.
\label{fig:fig8}}
\end{center}
\end{figure}

An experimental manifestation of the survival probability $S^2$ comes from comparing hard
diffraction at the Tevatron with that observed at HERA.  The relevant plot was presented by
Snow \cite{SNOW}, see also \cite{CDF}.  It shows the recent CDF measurements of
diffractive dijet production as a function of $\beta$, together with the expectations based on
convoluting the parton densities of the Pomeron (and those of the proton) with the
partonic-level cross sections of the hard subprocess.  $\beta$ is the momentum fraction of the
Pomeron entering the hard subprocess.  The parton densities of the Pomeron are determined
from diffractive deep-inelastic-scattering data at HERA \cite{MARAGE}.  The partonic
structure of the Tevatron dijet production and HERA diffractive DIS are sketched in Fig.~8.
However, as the CDF plot \cite{CDF,SNOW} shows, when the HERA Pomeron densities are used to
estimate Tevatron dijet production, the factorized prediction turns out to be about an order of
magnitude larger than the data.  A key assumption of the factorization estimate is that the
survival probability of the rapidity gap, associated with the Pomeron exchange, is the same in
Figs.~8(a) and 8(b).  Comparison of the diagrams shows that the breakdown of factorization
is an inevitable consequence of QCD, and occurs naturally, due to the small probability
$S^2$ for the rapidity gap to survive the soft rescattering which occurs between the incoming
hadrons at the Tevatron, but which is absent between the electron and the proton at HERA.
In fact, a detailed study of this comparison, including the $\beta$ dependence, tells us about
the dependence of the survival probability on the kinematic variables \cite{KKMR}, and
opens the door to the application to many other hard processes with rapidity gaps, including
those discussed by Khoze \cite{VAK}.

\section{The QCD Pomeron and small $x$ physics}

Lipatov gave a comprehensive survey of the Pomeron before and after QCD \cite{LIP}.
Since he was either close to, or pioneered, all these developments, it was a masterly and
informative summary.  The Pomeron before QCD was the subject of Section~4.  That
Pomeron was a Reggeon, even if it has a rather unusual Regge trajectory, which in the 
time-like ($t > 0$) region is mixed with the $f$ and $f^\prime$ meson trajectories 
\cite{KAID}.

In QCD the gluon and quark $t$-channel exchanges are themselves reggeized.  The QCD
Pomeron (or BFKL Pomeron as it is frequently called) is a compound state of two reggeized
gluons.  In general it may have multi-gluon components.  The BFKL framework allows the
behaviour of the scattering of hadronic objects with transverse scale $Q^2$ at centre-of-mass
energy $\sqrt{s}$ to be predicted in the domain $s \gg Q^2 \gg \Lambda_{\rm QCD}^2$.  In
the leading $\log$ (LL) approximation the cross section
\be
\label{eq:a18}
\sigma \; = \; \sum_{n = 0}^\infty \: c_n (\alpha_S \ln s)^n.
\ee
Since $\alpha_S \ln s \sim 1$ a resummation of the $\log$ terms is necessary.  BFKL carried
out this LL summation about 25 years ago with the result
\be
\label{eq:a19}
\sigma \; \sim \; s^\omega \quad\quad {\rm with} \quad\quad \omega \; = \; \bar{\alpha}_S \: 4
\ln 2,
\ee
where $\bar{\alpha}_S \equiv 3 \alpha_S/\pi$, see \cite{LIP}.  Hence we speak of the BFKL
or QCD Pomeron.  If $\bar{\alpha}_S \simeq 0.2$, then $\omega \simeq 0.5$.  This result
appears to be in contradiction with the observed rise of the relevant cross sections at large
$s$.  The data\footnote{For example, the behaviour of $F_2$ at small $x = Q^2/s$, or of
forward jets in deep-inelastic scattering at HERA, or of $\gamma^* \gamma^*$ scattering at
LEP, or jets separated by a large rapidity gap at the Tevatron.} indicate a power growth more
like 0.3 than 0.5.  It was expected that the resummation of the NLL corrections, that is of the $\alpha_S
(\alpha_S \ln s)^n$ terms, would remove the discrepancy.  Recently the computation of these terms has been completed \cite{LIP} and the corrections found to be large, giving \be
\label{eq:a20}
\omega \; = \; \bar{\alpha}_S \: 4 \ln 2 (1 - 6.3 \: \bar{\alpha}_S), \ee
which puts the usefulness of the whole perturbative approximation into question.  Fortunately
it was observed that a major part of remaining higher order corrections may be resummed to all orders \cite{RESUM}.  These all-order resummations bring the BFKL programme back under control.

At small Bjorken $x$ the gluon distribution, $f$, unintegrated over its transverse momentum
$k_t$, should exhibit BFKL behaviour.  That is a characteristic $f (x, k_t^2) \sim x^{-
\omega}$ growth as $x \rightarrow 0$, accompanied by diffusion in $k_t$.  The transverse
momenta are not ordered in the small $x$ BFKL evolution leading to a Gaussian-type form of
$f$ in $\ln k_t^2$ with a width which grows as $\sqrt{\ln 1/x}$ as $x \rightarrow 0$. 
Ultimately the diffusion will be a problem since it leads to increasingly important 
contributions from the infrared domain of $k_t^2$ where the BFKL equation
is not expected to be valid.

Stirling \cite{STIR} discussed ways of studying the BFKL Pomeron at the LHC.  He
concentrated on the original idea of Mueller and Navelet, that is to study the correlations
between two jets widely separated in rapidity.  It was argued that the best BFKL indicator is
the rate of the weakening of the azimuthal (back-to-back type) correlation between the jets, as
the rapidity interval increases --- a manifestation of the diffusion in $k_t$.  The Tevatron data
show much less decorrelation than predicted by naive BFKL, but a much more realistic
Monte Carlo has been developed \cite{JEPPE} which allows the proper constraints on phase
space to be imposed.  Predictions for the LHC were presented.

De Roeck \cite{ALB} discussed the opportunities at the LHC to observe the behaviour of
parton densities at very small $x$.  In particular he emphasized that it may be possible to
probe the gluon distribution in the $x \simeq 10^{-6}-10^{-5}$ and $Q^2 \simeq 5~{\rm
GeV}^2$ domain by observing either prompt photon production $(gq \rightarrow \gamma q)$
or Drell-Yan production at very large rapidities.  The latter process involves a convolution to
allow for the $g \rightarrow q\bar{q}$ transition, which is required for a gluon-initiated
reaction; consequently somewhat large values of the gluon $x$ are probed.  He pointed out
that this domain may allow the shadowing corrections to $xg (x, Q^2)$ to be studied.
Prompted by this talk, a quantitative study was performed using a unified evolution equation
which embodies both BFKL and DGLAP behaviour and which incorporates the leading $\ln
1/x$ triple-Pomeron vertex \cite{KIMB}.  The shadowing corrections were found to be small
in the HERA domain, but lead to about a factor of two suppression of the gluon in the $x \sim
10^{-6}, Q^2 \sim 4~{\rm GeV}^2$ region, which should be accessible in the experiments at
the LHC.

\section{QED $\ell^+ \ell^-$ production as an LHC luminosity monitor}

One possibility to measure the luminosity at the LHC is to observe exclusive lepton-pair
production via photon-photon fusion
\be
\label{eq:a21}
pp \; \rightarrow \; p \: + \: \ell^+ \ell^- \: + \: p,
\ee
where $\ell = e$ or $\mu$, see \cite{SHAM,CARON}.  The Born amplitude (Fig.~9(a)) may
be calculated within QED \cite{BGMS}, and there are no strong interactions involving the leptons in
the final state.  The main phenomenological questions concern, first, the size of the absorptive
corrections arising from inelastic proton-proton rescattering (sketched symbolically in
Fig.~9(b)) and, second, how to suppress the proton dissociation contributions of Fig.~9(c).
These questions are addressed in Ref.~\cite{KMOR}.  The dissociation contributions vanish
as $q_{it} \rightarrow 0$, due to gauge invariance, where the $q_i$ are defined in Fig.~9(a).  Since it is difficult to measure a leading proton with $q_t \lapproxeq 30$~MeV, it is proposed \cite{SHAM} to select events with very small transverse momentum of the lepton pair
\be
\label{eq:a22}
p_t (\ell^+ \ell^-) \; \equiv \; | \mbox{\boldmath $q$}_{\ell^+ t} \: + \: \mbox{\boldmath $q$}_{\ell^- t} |,
\ee
with typically $p_t (\ell^+ \ell^-) \lapproxeq 30$~MeV.  Moreover the rescattering correction of Fig.~9(b)
is suppressed because the main part of the Born amplitude (Fig.~9(a)) comes from large
impact parameters\footnote{Even more, the amplitude of Fig.~9(b) is greatly suppressed at
small $b$ by a $J_z = 0$ selection rule.} $b$, whereas the rescattering occurs at smaller $b$.

\begin{figure}[htb]
\begin{center}
\epsfig{figure=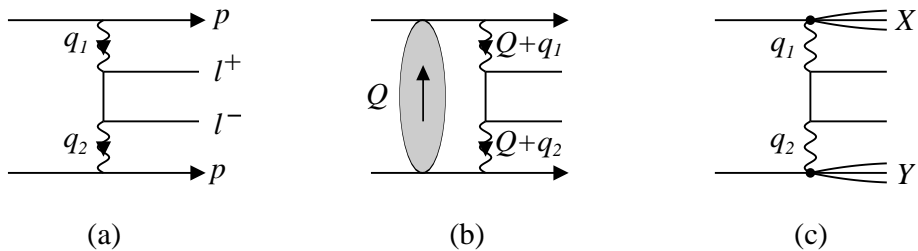,height=1.5in}
\caption{(a) lepton pair production which may be used as a luminosity monitor at the
LHC, (b) a typical rescattering correction, and (c) possible contamination coming from proton
dissociation into $X, Y$ systems.  Diagram (b) must not be viewed literally.  The Pomeron
exchanged between the protons should be viewed as a gluon ladder, and the dominant
contribution comes from four different configurations, corresponding to the photons being
emitted either before or inside the ladder.
\label{fig:fig9}}
\end{center}
\end{figure}

The detection of the $\mu^+ \mu^-$ and $e^+ e^-$ processes differ.  Hence their use as a
luminosity monitor is different.  First we discuss $\mu^+ \mu^-$ production
\cite{SHAM,CARON,KMOR}.  To identify muons (and separate them from $\pi^\pm$
mesons) they have to have rather large transverse energy, $E_t \gapproxeq 5$~GeV.  It is still
possible to satisfy the $p_t (\mu^+ \mu^-) < 30$~MeV cut, but the cross section is
significantly reduced.  The main contribution to the rescattering correction comes from the
$Q_t \approx p_t (\mu^+ \mu^-)$ domain, where $Q$ is the loop momentum in Fig.~9(b).
As a consequence the correction is given by $C \sigma_{\rm inel} p_t^2 (\mu^+ \mu^-)$, where $C$ is a known, small numerical coefficient \cite{KMOR}.  If $p_t (\mu^+ \mu^-) = 30$~MeV then the correction is
only 0.13\%.  In addition to the small $p_t (\mu^+ \mu^-)$ cut, it is proposed \cite{SHAM} to fit
the observed distribution in the muon acoplanarity angle $\phi$ in order to distinguish the
elastic mechanism, Fig.~9(a), via its prominent peak at $\phi = 0$, from the background
processes which are flat in $\phi$.  Although the $E_t$ and $p_t (\mu^+ \mu^-)$ cuts reduce the cross
section, the muons have the advantage that we may trace the tracks back to the interaction
vertex, and hence isolate the interaction in pile-up events.  So, in principle, $\mu^+ \mu^-$
production can act as a luminometer in high luminosity LHC runs.  An accuracy of $\pm
2\%$ is claimed provided the muon trigger is good enough \cite{SHAM,CARON}.

For $e^+ e^-$ production we do not need to select events with large $p_{et}$, and so we may
consider the small $p_{et}$ domain where the $e^+ e^-$ production cross section is much
larger, and where the rescattering correction becomes totally negligible.  This will require a
dedicated detector in the forward regions for electrons of energy about 5~GeV.  If the
threshold energy were reduced to 1~GeV then the signal is increased by 15 and the
signal-to-background ratio is substantially improved \cite{SHAM}.  It is claimed that an
absolute luminosity measurment down to $\pm 1\%$ is possible for low luminosities ${\cal
L} \lapproxeq 10^{32}~{\rm cm}^{-2}~{\rm s}^{-1}$, but for high luminosity the $e^+ e^-$
method may be limited by pile-up effects.

\section{$W$ and $Z$ production as a luminosity monitor}

$W$ and $Z$ production in high energy $pp$ collisions have clean signatures through their
leptonic decay modes, $W \rightarrow \ell \nu$ and $Z \rightarrow \ell^+ \ell^-$, and so may
be considered as potential luminosity monitors \cite{FD,CARON,KMOR}.  A vital
ingredient is the accuracy with which the cross sections for $W$ and $Z$ production can be
theoretically calculated.  The cross sections depend on parton distributions, especially quark
densities, in a kinematic region where they are believed to be reliably known.  Recent
determinations of $\sigma_{W,Z}$ at the LHC are shown in Fig.~10.  The solid squares and
triangles are from the NLO parton analyses of \cite{MRST1} and the final two predictions are
from the NLO and NNLO analyses of \cite{MRST2}.  The two major uncertainties appear to
be due to the value of $\alpha_S$ and to using different parton densities labelled by
$q\!\uparrow$ and $q\!\downarrow$.  The $\alpha_S \! \uparrow$ and $\alpha_S \!
\downarrow$ values correspond to changing $\alpha_S (M_Z^2)$ by $\pm 0.005$, which is
probably too conservative, so a $\pm 2\%$ uncertainty in $\sigma_{W,Z}$ is more realistic
from this source.  The normalisation of the input data used in the global parton analyses is
another source of uncertainty in $\sigma_{W,Z}$.  The HERA experiments provide almost
all of the data used in the global analyses in the relevant small $x$ domain.  The
$q\!\uparrow$ and $q\!\downarrow$ parton sets correspond to separate global fits in which
the HERA data have been renormalized by $\pm 2.5\%$ respectively.  Allowing for these
uncertainties, we conclude that the cross sections of $W$ and $Z$ production are known to
be about $\pm 4\%$ at the LHC energy.  For a precise measurement allowance should be
made for $W^+ W^-$ pair production and for $W$ bosons produced via $t$-quark decays,
which produce about $1\%$ of the total signal.

Caron \cite{CARON} discussed the experimental requirements of using $W$ and $Z$ production
to determine the luminosity at the LHC.

\begin{figure}[htb]
\begin{center}
\epsfig{figure=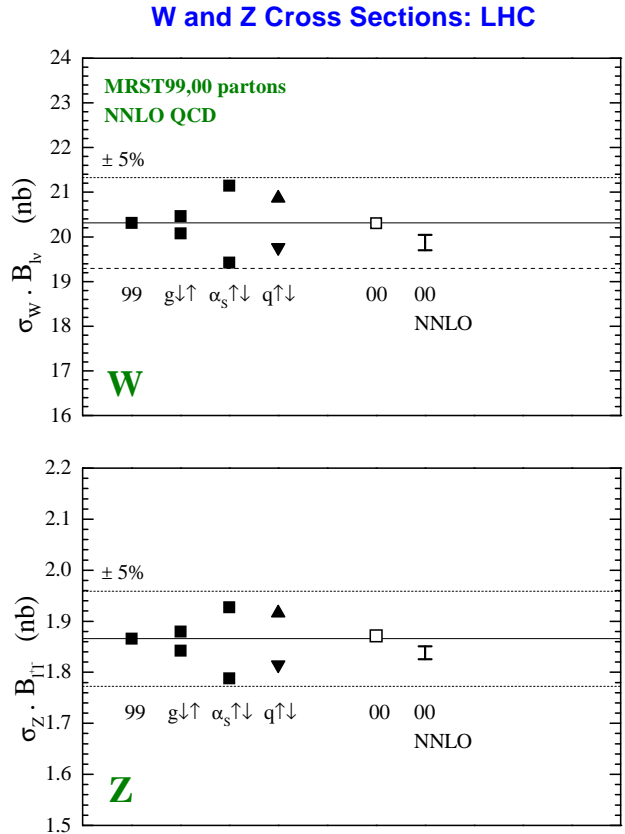,height=4.5in}
\caption{The predictions of the cross sections for $W$ and $Z$ production, and
leptonic decay, at the LHC \cite{MRST1,MRST2}.
\label{fig:fig10}}
\end{center}
\end{figure}

\section{Summary on luminosity determination}

The luminosity determinations based on the measurement of the forward elastic cross section
and on two-photon $e^+ e^-$ production can only be made in low luminosity runs, and
require dedicated forward detectors and triggers.  On the other hand, the measurement of $W$
or $Z$ and two-photon $\mu^+ \mu^-$ production may be performed at high luminosity with
the central detector.  The latter process requires a good di-muon trigger with thresholds of $E_t \lapproxeq 5$~GeV for each muon.

In principle, we may monitor the luminosity using any process, with a significant cross
section, which is straightforward to detect cleanly.  For example, it could be single-pion
production in a given rapidity and $p_t$ domain or inclusive $\mu^+ \mu^-$ production in a
well defined kinematic region.  In this way we may determine the relative luminosity and
calibrate the \lq\lq monitor\rq\rq\ by comparing the number of events detected for the \lq\lq
monitor\rq\rq\ reaction with the number of events for a process whose cross section is known.
This has the advantage that the calibration may be carried out in a low luminosity run.

The desired goal of measuring the luminosity to better than $\pm 5\%$ definitely seems
attainable.  We note that for applications where it is sufficient to know the parton-parton
luminosity, better accuracy can be achieved.

\section{Final Observations}

One view of forward or soft physics, which dominates the interactions at the LHC, is that it is
an unfortunate unavoidable complication to the exciting rare events which we hope to see.  It
may be useful as a luminosity monitor, but the Pomeron is boring.

Another view expressed at the meeting\footnote{This viewpoint is also always emphasized by Bjorken \cite{BJ1,BJ2}, and resulted in the FELIX proposal \cite{FELIX}.} is that to avoid the Pomeron is to deny our birthright.  Almost all QCD is contained in the Pomeron.  The Pomeron indirectly spawned string theory
and sent a galaxy of physicists spinning off into higher dimensions.  Most of the rich HERA
physics is driven by the Pomeron.  The \lq soft\rq\ Pomeron in the non-perturbative domain
has fascinating Regge properties, leading to glueballs and mixing with $q\bar{q}$ states.
With the advent of QCD we were reminded that the quark and gluon are not elementary, but
Reggeons --- and that the \lq hard\rq\ or QCD Pomeron is the compound state of two (or
more) Reggeized gluons.  The correct interpolation between the hard and the soft regimes is
an exciting fundamental problem yet to be solved.

This meeting has offered the opportunity of a wider perspective of these two viewpoints.  Moreover,
we were able to glimpse the great potential, and the experimental challenges, of the LHC.

It is a special pleasure to thank, on behalf of all of the participants, Dan-Olof Riska, Risto Orava and the other members of the organizing committee, together with Laura Salmi, for arranging such an excellent Workshop, and for their hospitality in Helsinki.  I gratefully acknowledge the help that I have received from Aliosha Kaidalov, Valery Khoze, Misha Ryskin and Stefan Tapprogge on the subject of this Workshop.

\end{document}